\documentclass[a4paper,fleqn,numbers]{cas-dc}

\usepackage{etoolbox,soul}
\apptocmd{\sloppy}{\hbadness 10000\relax}{}{}
\usepackage[numbers]{natbib}

\def\tsc#1{\csdef{#1}{\textsc{\lowercase{#1}}\xspace}}
\tsc{WGM}
\tsc{QE}

\begin{document}

\let\WriteBookmarks\relax
\def\floatpagepagefraction{1}
\def\textpagefraction{.001}

\shorttitle{Estimating the Contribution of Galactic Neutrino Sources}


\title [mode=title]{Estimating the Contribution of Galactic Neutrino Sources} 
\author[1]{Mohadeseh {Ozlati Moghadam}}[orcid=0009-0003-2479-1863]
\ead{mohadeseh.ozlati@uni-potsdam.de}

\author[1]{Kathrin Egberts}
\ead{kathrin.egberts@uni-potsdam.de}

\author[1]{Rowan Batzofin}

\author[1]{Constantin Steppa}

\author[2]{Elisa Bernardini}

\affiliation[1]{organization={University of Potsdam},
            addressline={Karl-Liebknecht-Str.\,24-25}, 
            postcode={14476},
            city={Potsdam},
            country={Germany}}
            
\affiliation[2]{organization={Università degli Studi di Padova},
            addressline={Via Marzolo}, 
            postcode={8 - 35131},
            city={Padova},
            country={Italy}}

\shortauthors{M.O. Moghadam et al.} 

\begin{abstract}
The Milky Way hosts astrophysical accelerators capable of producing high-energy cosmic rays.
These cosmic rays can interact with the interstellar medium (ISM) across the Galaxy to produce neutrinos and gamma rays (propagation component), while their interactions with ambient material at their acceleration sites, such as supernova remnants, can give rise to the source component of the gamma-ray and neutrino flux. 
In this paper, 
we estimate the source component of the Galactic neutrino flux using simulated populations of Galactic gamma-ray sources. We compare our results with observations from neutrino experiments in the energy range of 1–30 TeV.  
Using simulated populations of Galactic TeV gamma-ray sources, we exploit the correlation between gamma rays and neutrinos 
and introduce a bracketing approach to constrain
the range for the source contribution of the Galactic neutrino flux. For the upper limit, we used a simulation describing the entity of Galactic gamma-ray sources, whereas the lower limit was estimated using the hadronic component of the Galactic supernova remnant population. Our results show that the difference between this maximum and minimum is less than an order of magnitude and the flux range is comparable to the Galactic neutrino flux from the cosmic-ray interaction with the ISM. 
The results agree with the observed signals from IceCube and ANTARES and suggest that the propagation component, combined with the minimum source contribution predicted by the supernova-remnant model, approaches the observed neutrino flux, leaving little room for significant enhancements of the emission originating from propagating cosmic rays.

\end{abstract}

\makeatletter\def\Hy@Warning#1{}\makeatother
\maketitle
\section{Introduction}
\sloppy
Galactic sources can accelerate cosmic rays (CR) up to 
PeV energies.
When these high-energy hadrons interact with the interstellar medium (pp interaction), they can generate neutral ($\pi^0$) and charged ($\pi^\pm$) pions, which subsequently decay into gamma rays and neutrinos. Neutrinos trace hadronic processes, while gamma rays may also result from leptonic mechanisms such as bremsstrahlung and inverse Compton scattering.
Identifying whether gamma rays are leptonic or hadronic in origin is a challenge for gamma-ray observations. Therefore, neutrino measurement can play a crucial role in this distinction. \\
The Milky Way has been extensively studied in gamma rays, revealing CR accelerators such as supernova remnants (SNRs), pulsar wind nebulae (PWNe), microquasars, and others \cite{HGPS}, as well as the traces of CR propagation in the form of diffuse gamma-ray emission \cite{Fermi, Abramowski_2014, Cao_2023}.\\
Neutrino astronomy is now opening new frontiers for exploring our Galaxy, with recent advancements confirming the presence of Galactic neutrinos. The Galactic neutrino flux can arise either from the interaction of CRs with the interstellar medium (propagation component) or from CR interactions occurring at individual acceleration sites, such as supernova remnants (source component).\\
IceCube has detected neutrinos from the Galactic plane with a significance level of 4.5$\sigma$ \cite{IceCubeneutrino} using templates derived from gamma-ray observations. 
Similarly, neutrinos from the Galactic Ridge region ($|l| < 30^\circ, |b| < 2^\circ$) have been reported by the ANTARES Collaboration with a significance level of 2.2$\sigma$ \cite{Galactic_ridge}. \\
This measured diffuse neutrino flux can be expected to be a combination of the propagation component and the component of unresolved sources, with unknown relative contributions because
current statistics and angular resolution are not sufficient to distinguish between the two.\\
The interpretation of the Galactic neutrino flux inevitably relies on the vast dataset of gamma-ray data available on the Galactic plane. 
The gamma-ray/neutrino connection has been intensively studied to interpret the IceCube and ANTARES measurements, following different approaches. One approach uses direct measurements of gamma rays to derive neutrino fluxes, either for the diffuse gamma-ray emission, e.g. with Fermi-LAT, or by integrating over the fluxes of measured gamma-ray sources in the TeV to PeV range \cite{icecube_galactic_ridge, desert_Halzen, neutrino_Gagliardini,2023ApJ...957L...6F}. \\
Alternatively, gamma-ray data are used to tune the modeling of the propagation component or to constrain the modeling of populations of individual source types, which are then used to determine their cumulative neutrino flux including unresolved sources \cite{Vecchiotti_IceCube, ANTARES_Vecchiotti_2023, Constrains_origin_neutrino, 2024PhRvD.109d3007A, 2025PhRvD.111f3035L, 2025arXiv250218268D}.\\
One of the most common methods for correlating the gamma-ray flux to the neutrino flux is based on the proportional relationship between the two, as outlined by \cite{ahlera_2014}:
\begin{equation}
   \textstyle E_\gamma \varphi_\gamma\left(E_\gamma\right) \approx e^{\frac{-d}{\lambda \gamma \gamma}} \frac{1}{3} \sum_{\nu_l} E_\nu \varphi_{\nu_l}\left(E_\nu\right)\,
   \label{common_methof}
\end{equation}
where $\varphi_{\nu_{\alpha}}(E_{\nu})$ represent the differential flux, $E_{\nu}$  is the energy of neutrino with flavour $l$, $\varphi_{\gamma}(E_{\gamma})$ represents the gamma-ray differential flux and $E_{\gamma}$ is the gamma-ray energy. The exponential describes gamma absorption ($d$ is the distance to the source, $\lambda_{\gamma\gamma}(E)$ the gamma-ray interaction length). \\
The gamma-ray energy can be simply approximated as $E_{\gamma} \approx 2E_{\nu}$. Since we only consider Galactic sources, we assume that the absorption of gamma-rays is negligible ($\frac{-d}{\lambda \gamma \gamma}\approx 0$) and the flux from each neutrino flavor is the same due to oscillations. Equation \ref{common_methof} then translates into \cite{join_Hawc_IceCube}: 
\begin{equation}
     \varphi_{\nu_{\mu}}(E_{\nu_{\mu}}) \approx 2 \varphi_\gamma(2E_{\nu_{\mu}})  
     \label{eq2}
\end{equation}
More precise calculations require the parameterization of the pp interaction cross section to derive the resulting gamma-ray and neutrino fluxes \cite{Kelner}.

In this study, we provide predictions of the source contribution to the Galactic neutrino flux based on simulated gamma-ray populations. Rather than using individual sources or gamma-ray measurements \cite{icecube_galactic_ridge, desert_Halzen, neutrino_Gagliardini,2023ApJ...957L...6F} we rely on published models for the gamma-ray source populations, similar to \cite{Vecchiotti_IceCube, ANTARES_Vecchiotti_2023, Constrains_origin_neutrino, 2024PhRvD.109d3007A, 2025PhRvD.111f3035L, 2025arXiv250218268D}. 
Our work advances previous studies by introducing a bracketing approach that does not rely on a single assumed source population or observations of individual sources, but instead derives a robust envelope for the Galactic source contribution. \\
We estimate a maximum and a minimum neutrino flux for Galactic sources: For the maximum source contribution (model I), we used the entire population of Galactic TeV sources, derived and verified in \cite{Steppa_2020}, which includes also strong leptonic emitters such as PWNe, assuming their emission to be entirely hadronic. For the minimum scenario (model II), taken from \cite{batzofin2024SN}, we used the population of Galactic SNRs and calculated the neutrino flux only from the hadronic component of the gamma-ray emission.
For the correlation of the gamma-ray flux to the neutrino flux we use the Kelner parametrisation \cite{Kelner} in a simplified form to reduce computational costs.
 We then compare the maximum and minimum source flux with a calculation of the neutrino emission arising from CRs interacting with the ISM (propagation component).
\begin{figure}[t]
    \centering
    \includegraphics[width=\columnwidth]{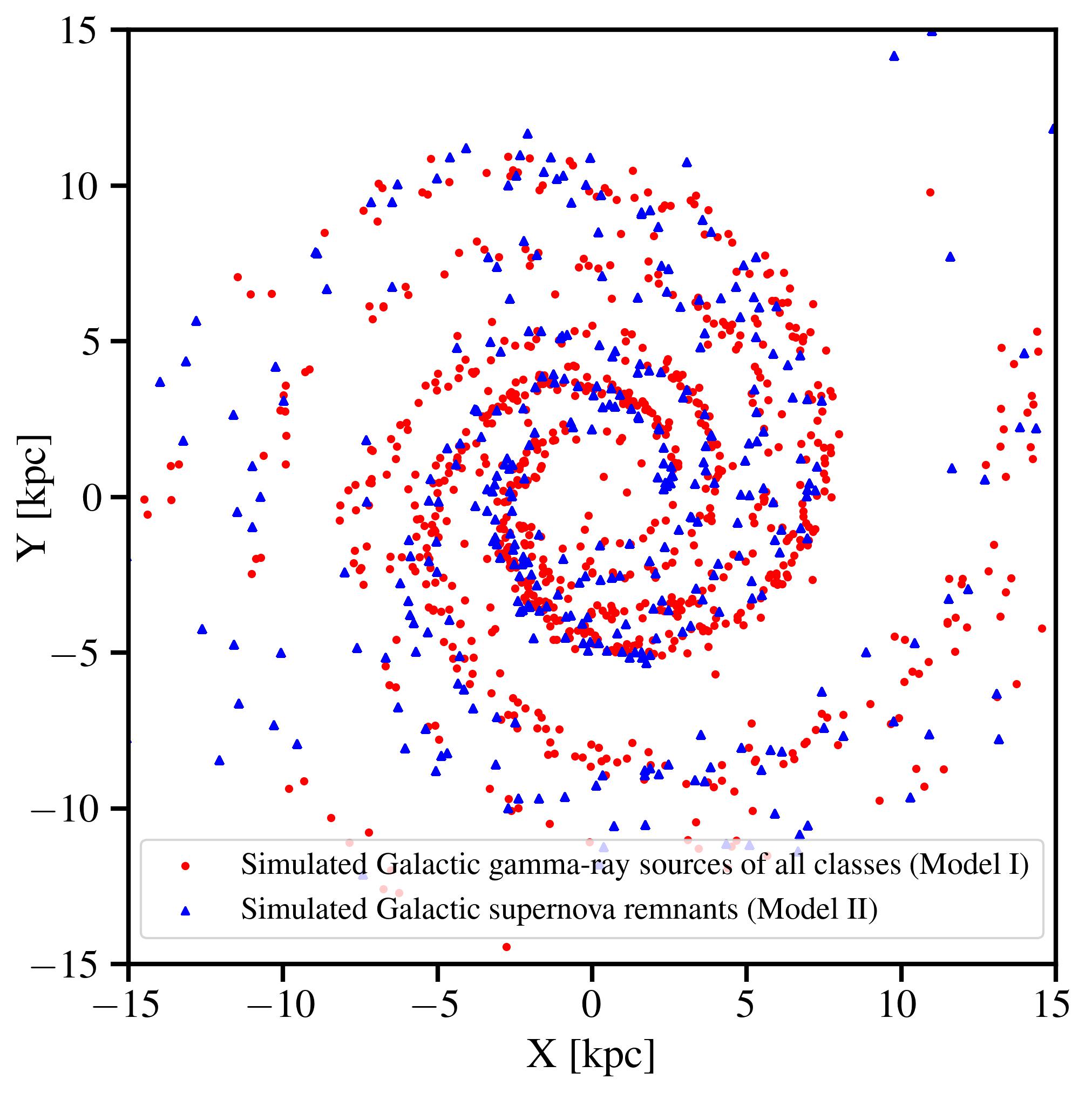}
  \caption{Spatial distribution of gamma-ray sources from a selected population of model $\mathrm{I}$ \cite{Steppa_2020} (red circles) and model $\mathrm{II}$ \cite{batzofin2024SN} (blue triangles). The sources follow a four-arm spiral spatial model in the Galactic plane.}
  \label{Spatial_dist}
\end{figure}
\section{Methodology}
\sloppy
We bracketed the Galactic neutrino flux with two simulated gamma‑ray source populations: \textbf{model I}, taken from \cite{Steppa_2020}, serves as the maximum since it includes all different type of gamma ray sources and \textbf{model II}, taken from \cite{batzofin2024SN}, as the minimum estimate of the Galactic neutrino source contribution as it only includes the hadronic component of 
Galactic SNRs.\\
For both source models, their spatial distribution follows the matter distribution in the Galaxy as measured by Steiman-Cameron et al. \cite{Steiman-Cameron} and is illustrated in Fig.~\ref{Spatial_dist}.\\
Large numbers of populations were simulated for each model and the resulting neutrino emission was averaged over these populations. 
Each population in model I \cite{Steppa_2020} 
describes different classes of very high-energy (VHE) Galactic gamma-ray sources incorporating both simulated sources and observed ones from the H.E.S.S. Galactic Plane Survey (HGPS) \cite{HGPS}. 
The luminosities and radii of these sources are independently sampled from power-law distributions, which have been fitted in the subsample of detectable sources to the distributions observed in the HGPS dataset.\\
The populations in model II represent Galactic supernova remnants (SNRs). This model \cite{batzofin2024SN} is based on the supernova remnant evolution and calculates the gamma-ray emission based on an assumed acceleration efficiency, electron-to-proton ratio, and cosmic-ray spectral index. The free parameters are optimised to achieve the best agreement with the HGPS data.
The optimal set of parameters used for our simulations produce 97$\%$ of simulations in agreement with the HGPS and consist of:
cosmic-ray spectral index of 4.2 (in momentum space), an electron-to-proton ratio $K_{\rm ep} = 10^{-5}$, an acceleration efficiency of $\eta$ = 9\%, and the Sedov-Taylor phase lasting $T_{\rm ST}$ = 20 kyr. The maximum energy of protons was estimated using the Bell description \cite{Bell} based on the maximum non-resonant hybrid growth rate \cite{batzofin2024SN}. 

To establish a connection between gamma-ray sources and their corresponding neutrino emissions, we assume absorption effects to be negligible and consider the neutrino flavour ratio as $\nu_{\mu},\nu_{\tau},\nu_{e} = 1:1:1$, which is a direct consequence of neutrino oscillations.\\
We based our neutrino flux estimate on the work of Batzofin and Komin \cite{Rowan}. 
Their study used Kelner's parameterisation \cite{Kelner} to calculate the neutrino flux for all HGPS sources excluding identified PWNe, under the assumption that gamma rays are purely hadronic. \\
Using these results, we independently normalized the total simulated gamma-ray flux for each model to the total observed H.E.S.S. gamma-ray flux reported in \cite{Rowan} and calculated a corresponding scaling factor for each model. Namely:
\begin{equation}
\alpha \;=\;
\frac{\displaystyle
        \frac{1}{N_{\mathrm{pop}}}\,
        \sum_{j=1}^{N_{\mathrm{pop}}}
        \Phi_{\gamma,j}^{\mathrm{sim}}}
     {\displaystyle
        \sum_{k=1}^{N_{\mathrm{obs}}}
        \int_{1\ \mathrm{TeV}}^{100\ \mathrm{TeV}}
        \phi_{\gamma,k}^{\mathrm{H.E.S.S.}}(E)\,dE},
\label{eq:alpha}
\end{equation}
where
\begin{equation}
    \Phi_{\gamma,j}^{\mathrm{sim}}
\;=\;\sum_{s=1}^{n_j}\int_{1\ \mathrm{TeV}}^{100\ \mathrm{TeV}}\phi_{\gamma,js}^{\mathrm{sim}}(E) \,dE
\end{equation}
is sum of the integrated \(\gamma\)-ray flux of all the sources (\(n_j\)) in each population (cm\(^{-2}\) s\(^{-1}\)).  
\(N_{\mathrm{pop}}\) is the total number of simulated populations in each model;  
\(N_{\mathrm{obs}}\) is the number of H.E.S.S.\ sources included in the Batzofin calculation \cite{Rowan};  
\(\phi_{\gamma,k}^{\mathrm{H.E.S.S.}}(E)\) is the differential spectrum of \(k\)th H.E.S.S. source \ in the 1–100 TeV band (the choice of this energy range is motivated by the integration interval used in \cite{Steppa_2020}).\\

We multiplied the scaling factor by the total neutrino flux that Batzofin et al. \cite{Rowan} predict for the H.E.S.S. source catalogue, thereby obtaining the maximum- and minimum-case Galactic neutrino flux:
\begin{equation}
\phi_{\nu}^{\text{max/min}}(E)
  \;=\;
  \alpha_{\text{max/min}}\,
  \phi_{\nu, sum}^{\mathrm{\text{Predicted for H.E.S.S}}}(E)
\end{equation}
where $\Phi_{\nu}^{\mathrm{HESS}}(E)$ is the summed neutrino spectrum predicted for all H.E.S.S.\ sources given in Ref.~\cite{Rowan}, and $\alpha_{\text{max}}$ and $\alpha_{\text{min}}$ are the scaling factors obtained for model I and II, respectively.\\
\begin{figure}[t]
  \includegraphics[width=\columnwidth]{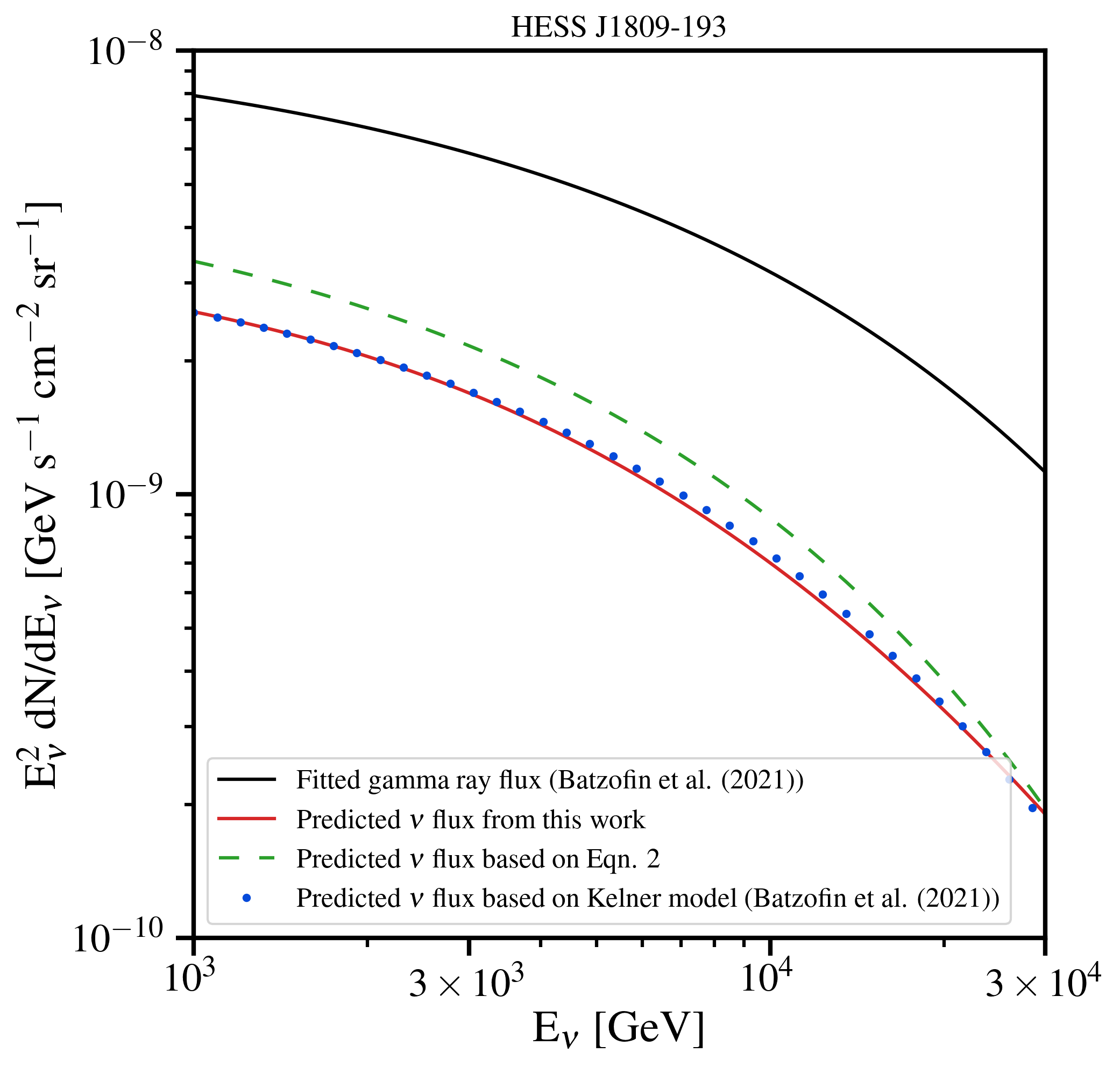}
  \caption{Comparison of neutrino flux estimates for a single H.E.S.S. source. The solid orange line represents the neutrino flux calculated using the method described in this work; the purple dashed line shows the neutrino flux obtained from the common proportional gamma–neutrino flux relationship (Equation \ref{eq2}); and the green dotted line indicates the flux based on the Kelner model from \cite{Rowan}. The fitted gamma-ray flux is included for reference (gray solid line at the top).}
    \label{one_source}
\end{figure}
Using the same method, but only including the sources in the Galactic Ridge region ($|l|< 30^\circ, |b|< 2^\circ$), we calculate the neutrino flux from the Galactic Ridge as well. \\
This method allows for a direct estimate of the neutrino flux based on a given gamma-ray flux, without the need for detailed and computationally expensive calculations of the proton distribution, as required by the Kelner parametrisation \cite{Kelner}. It should be noted that a direct consequence of this procedure is that our source spectra inherits a cutoff, which originates from the observed H.E.S.S. gamma-ray sources in \cite{Rowan} and is passed on to the neutrino spectrum. Consequently, the predictive power of the model decreases above $10$ TeV together with the H.E.S.S. sensitivity.

Fig.~\ref{one_source} shows the validation of this approach for a single source (HESS J1809-193). Together with the gamma-ray flux of the source, the neutrino flux predictions using our method are compared with the one obtained using the Kelner \cite{Rowan} and the approximation using Eqn.~\ref{eq2}. \\
Our estimated neutrino flux shows better agreement with the Kelner model compared to the simple gamma-neutrino correlation of Eqn.~\ref{eq2} up to energies of few tens of TeV.
We limit the energy range of our investigation to $<30$~TeV, reflecting the core sensitivity range of the H.E.S.S. experiment, which is used for the derivation of the conversion between gammas and neutrinos and the tuning of the population models.
\section{Results}
\begin{figure}[]
  \includegraphics[width=\columnwidth]{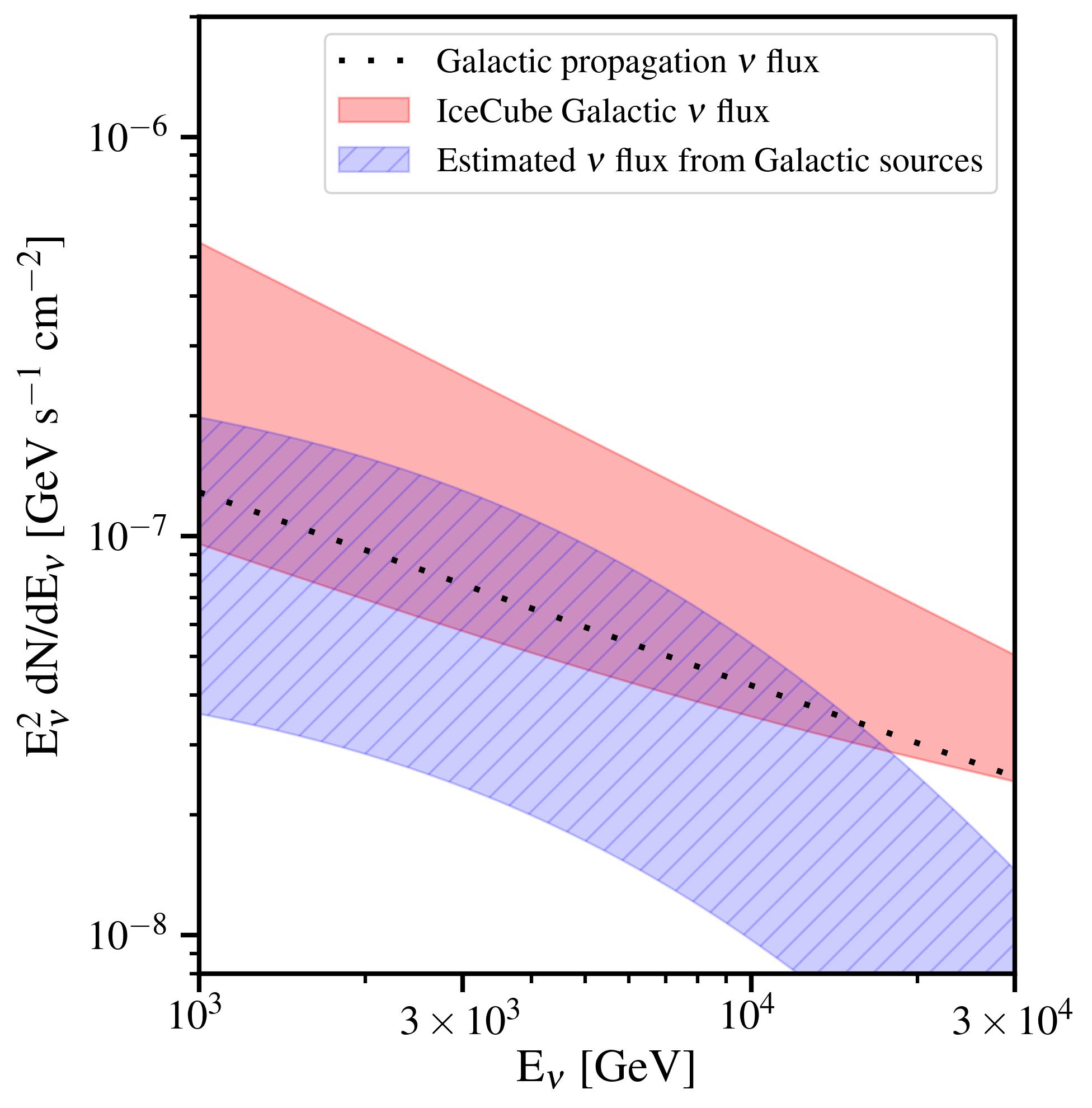}
  \caption{The source contribution to the Galactic neutrino flux per flavour, compared with IceCube data. 
  Results are based on all-sky (4$\pi$) templates and are
    presented as an all-sky flux.
    The red-shaded band encloses the three IceCube best‑fit normalisations obtained for different Galactic templates \cite{IceCubeneutrino}.  
    The blue-hatched band shows the flux from Galactic sources bracketed by model $\mathrm{I}$ (upper side: full VHE gamma‑ray population, assumed fully hadronic) and model $\mathrm{II}$ (lower side: hadronic component of the SNR population only).   
    The dotted line represents the propagation component calculated for a homogeneous CR flux according to equation \ref{diffuse_flux}. 
}
  \label{total_flux_IceCube}
\end{figure}

\begin{figure}[]
  \centering
  \includegraphics[width=\columnwidth]{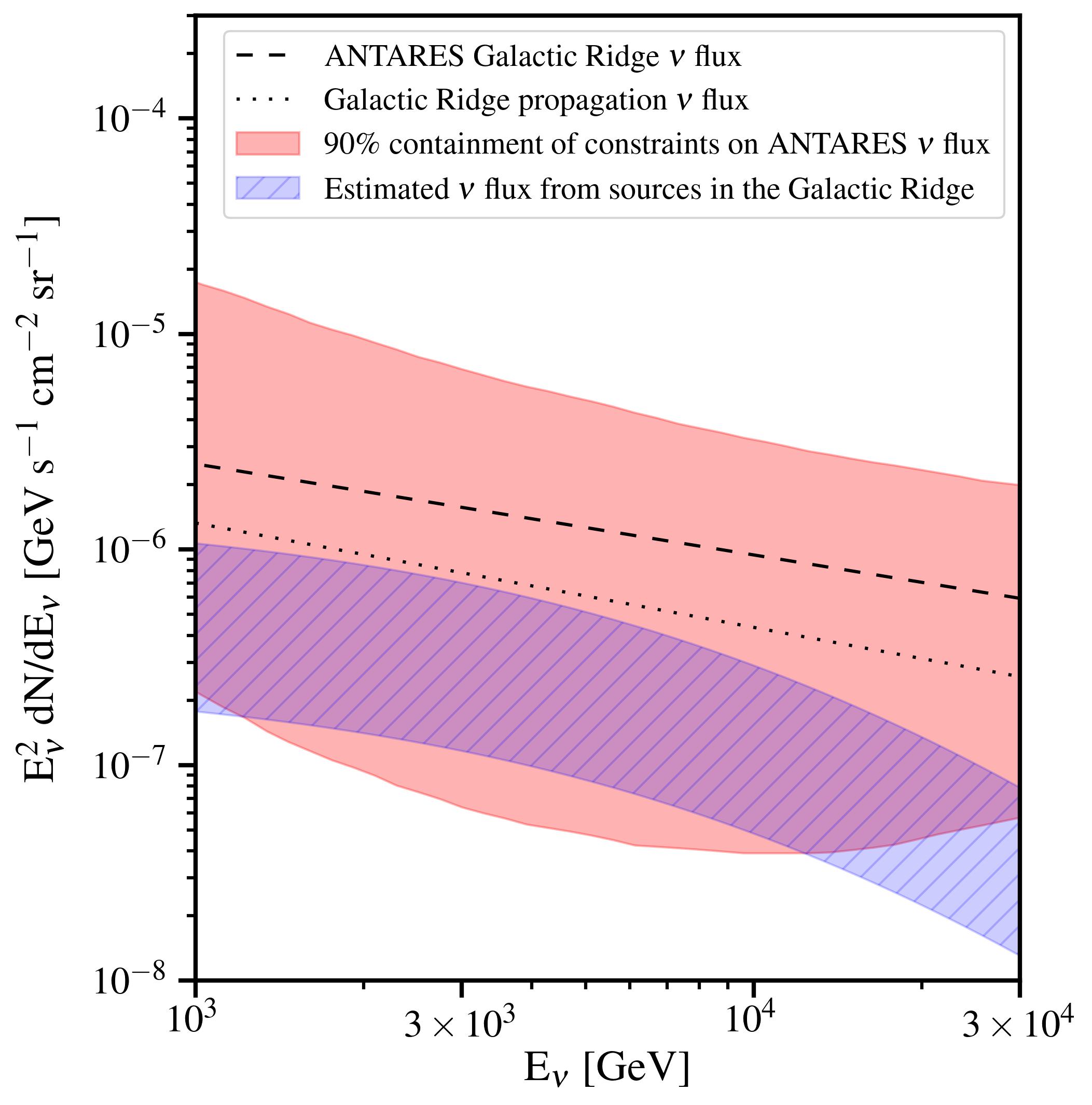}
  \caption{The source contribution to the neutrino flux per flavour in the Galactic Ridge region  
    ($-30^{\circ} < l < 30^{\circ}$, $-2^{\circ} < b < 2^{\circ}$), compared with ANTARES data \cite{Galactic_ridge}.  
    The blue-hatched band shows the flux from Galactic sources in that area, bracketed by model $\mathrm{I}$ (upper side: full VHE gamma‑ray population, assumed fully hadronic) and model $\mathrm{II}$ (lower side: hadronic component of the SNR population only).  
    The dotted curve shows the emission stemming from propagating CRs calculated for a homogeneous CR sea according to equation \ref{diffuse_flux}.  
    The red-shaded band denotes the ANTARES measurement with its $1\sigma$ uncertainty.
}
  \label{galactic_ridge}
\end{figure}

The resulting source contributions to the Galactic neutrino flux, along with the IceCube Galactic neutrino flux measurement \cite{IceCubeneutrino} and the ANTARES Galactic Ridge neutrino emission measurement \cite{Galactic_ridge} are shown in Fig.~\ref{total_flux_IceCube} and Fig.~\ref{galactic_ridge}, respectively.\\
The range of neutrino emission originating from Galactic sources is visualized by a band, constrained by an upper limit (model I) and a lower limit (model II). The difference between these maximum and minimum fluxes of the source contribution is less than an order of magnitude. For both the Galactic neutrino flux as well as the neutrino flux from the Galactic ridge, the neutrino emission from SNRs only (model II) cannot convincingly explain the data. Even the upper limit model I provides an estimate that lies below the IceCube and ANTARES measurements, although consistent within the errors below energies of 10 TeV.

We also compare our results with a minimum
propagation component of the neutrino flux, determined by \cite{Pagliaroli}:
\begin{multline}
\label{diffuse_flux}
\varphi_{\nu, \operatorname{diff}}\left(E_\nu, \hat{n}_\nu\right) = 
\frac{1}{3} \sum_{l=e, \mu, \tau} \Bigg[ 
\int_{E_\nu}^{\infty} dE \, 
\frac{d \sigma_l\left(E, E_\nu\right)}{d E_\nu} \\
+ \int_0^{\infty} dl \, 
\varphi_{CR}\left(E, r_{\odot} + l \hat{n}_\nu\right) 
n_{\mathrm{H}}\left(r_{\odot} + l \hat{n}_\nu\right) 
\Bigg]\,
\end{multline}
where $E_{\nu}$ and $\hat{n}_{\nu}$ are the neutrino energy and arrival direction and $\frac{d \sigma_l}{d E_\nu}$ is the differential cross section for the production of neutrinos and antineutrinos with flavour $l$ by a nucleon energy of $E$ in proton-proton interaction.
The term $\phi_{CR}(E,r)$ refers to the differential CR flux, $r_{\odot}=8.5$ kpc is the position of the Sun, and $n_{\mathrm{H}}$ is the gas density distribution function.\\
 We assumed a homogeneous CR flux throughout the Milky Way ($\phi_{CR}(E,r)=\phi_{CR,\odot}(E)$), a local CR spectrum parameterized by \cite{CR_parameteriation}, and used the gas distribution model from \cite{Shibata_2010}. The assumption of a constant CR flux makes this a lower limit estimate. For possible enhancements of the CR flux see Section III, Cases B and C of \cite{Pagliaroli}.

As shown in Fig.~\ref{total_flux_IceCube} (all sky) and Fig.~\ref{galactic_ridge} (Galactic ridge), 
the propagation and source contributions of the Galactic neutrino flux at energies below 10 TeV (where the cutoff in the sources becomes pronounced) are at a comparable level, with a slightly lower relative source contribution in the Galactic ridge. 
It has to be noted, however, that in the calculation of the propagation component there are still significant possibilities for enhancing the flux, e.g. via unaccounted gas in the matter distribution, nuclear enhancement factors, or an increase in the CR density beyond the locally measured level. \\
The neutrino measurements both by IceCube and ANTARES in the TeV range leave little room for 
such enhancements of the propagation component beyond the minimum flux 
assuming a homogeneous level of CRs in the Galaxy, if we take into account the requirement of the sum of propagation component and minimum source contribution of model II to be smaller than the measurement. 
\section{Summary and Conclusion}

In this paper, we present an estimate of the total Galactic neutrino flux, which includes contributions from both Galactic sources and emission stemming from CR propagation. \\
We introduce a bracketing approach for the source component that includes an upper limit based on the entity of VHE Galactic gamma-ray sources from \cite{Steppa_2020}, and a lower limit using the hadronic emission in SNRs from \cite{batzofin2024SN}. 
A minimum propagation component is calculated using a homogeneous CR distribution in the Galaxy according to equation \ref{diffuse_flux}.
Both source and the minimum propagation components are shown to be comparable, as illustrated in Fig.~\ref{total_flux_IceCube} (all sky) and Fig.~\ref{galactic_ridge} (Galactic ridge).\\
In comparison with IceCube and ANTARES results, our estimates demonstrate that the margins for CR enhancements in the Galaxy are limited. This demonstrates that the current state of neutrino measurements can provide valuable constraints. 
The significant remaining uncertainties in both the available neutrino data and the underlying models limit the predictive power of our conclusions and highlight the need for more precise measurements. \\
Observations from next-generation neutrino observatories such as KM3NeT and IceCube-Gen2 will be crucial in further constraining the Galactic neutrino flux and understanding its underlying mechanisms \cite{Bustamante_2023}.

\section*{Acknowledgment}
This work is funded by the Deutsche Forschungsgemeinschaft (DFG, German Research Foundation) with the grant 500120112. I also would like to thank Dr. Tiffany Collins for her valuable help in proofreading my paper and providing insightful suggestions that greatly improved its clarity and quality.

\printcredits

\bibliographystyle{cas-model1-num-names}
%
\bibliography{cas-refs_new}

@article{Galactic_ridge,
  title={Hint for a TeV neutrino emission from the Galactic Ridge with ANTARES},
  author={Albert, Arthur and Alves, S and Andre, Michel and Ardid, M and Ardid, S and Aubert, J-J and Aublin, J and Baret, Bruny and Basa, S and Becherini, Y and others},
  journal={Physics Letters B},
  volume={841},
  pages={137951},
  year={2023},
  publisher={Elsevier}
}

@article{Constrains_origin_neutrino,
  title={Constraints on the Origins of the Galactic Neutrino Flux Detected by IceCube},
  author={Desai, Abhishek and Vandenbroucke, Justin and Anandagoda, Samalka and Thwaites, Jessie and Romfoe, MJ},
  journal={The Astrophysical Journal},
  volume={966},
  number={1},
  pages={23},
  year={2024},
  publisher={IOP Publishing}
}

@article{IceCubeneutrino,
  title={Observation of high-energy neutrinos from the Galactic plane},
  author={IceCube Collaboration and Abbasi, R and Ackermann, M and Adams, J and Aguilar, JA and Ahlers, M and Ahrens, M and Alameddine, JM and Alves Jr, AA and Amin, NM and others},
  journal={Science},
  volume={380},
  number={6652},
  pages={1338--1343},
  year={2023},
  publisher={American Association for the Advancement of Science}
}

@ARTICLE{HGPS,
       author = {{H.~E.~S.~S. Collaboration} and {Abdalla} and H. and {Abramowski} et al.},
        title = "{The H.E.S.S. Galactic plane survey}",
      journal = {Astronomy \& Astrophysics},
         year = 2018,
        month = apr,
       volume = {612},
          eid = {A1},
        pages = {A1},
          doi = {10.1051/0004-6361/201732098},
archivePrefix = {arXiv},
       eprint = {1804.02432},
 primaryClass = {astro-ph.HE},
       adsurl = {https://ui.adsabs.harvard.edu/abs/2018A&A...612A...1H}
}

@article{Kelner,
  title={Energy spectra of gamma rays, electrons, and neutrinos produced at proton-proton interactions in the very high energy regime},
  author={Kelner, Stanislav R and Aharonian, Felex A and Bugayov, Vistcheslav V},
  journal={Physical Review D—Particles, Fields, Gravitation, and Cosmology},
  volume={74},
  number={3},
  pages={034018},
  year={2006},
  publisher={APS}
}

@article{Rowan,
  title={Predicting the Neutrino Emission for the Sources in the HESS Galactic Plane Survey},
  author={Batzofin, Rowan and Komin, Nukri},
  journal={Proceedings of Science, ICRC (online)},
  year={2021}
}

@ARTICLE{Fermi,
       author = {{LAT Collaboration} and {Ackermann}, M. and {Ajello} et al.},
        title = "{Fermi-LAT Observations of the Diffuse {\ensuremath{\gamma}}-Ray Emission: Implications for Cosmic Rays and the Interstellar Medium}",
      URL = {https://in2p3.hal.science/in2p3-00685261},
  JOURNAL = {{The Astrophysical Journal}},
  PUBLISHER = {{American Astronomical Society}},
  VOLUME = {750},
  PAGES = {3},
  YEAR = {2012},
  DOI = {10.1088/0004-637X/750/1/3},
  HAL_ID = {in2p3-00685261},
  HAL_VERSION = {v1},
}

@article{Steppa_2020,
  title={Modelling the Galactic very-high-energy $\gamma$-ray source population},
  author={Steppa, Constantin and Egberts, Kathrin},
  journal={Astronomy \& Astrophysics},
  volume={643},
  pages={A137},
  year={2020},
  publisher={EDP Sciences}
}

@article{batzofin2024SN,
  title={The population of Galactic supernova remnants in the TeV range},
  author={Batzofin, Rowan and Cristofari, Pierre and Egberts, Kathrin and Steppa, Constantin and Meyer, Dominique M-A},
  journal={Astronomy \& Astrophysics},
  volume={687},
  pages={A279},
  year={2024},
  publisher={EDP Sciences}
}

@article{icecube_galactic_ridge,
  title={Hadronic nature of high-energy emission from the Galactic ridge},
  author={Neronov, Andrii and Semikoz, D and Aublin, J and Lamoureux, M and Kouchner, A},
  journal={Physical Review D},
  volume={108},
  number={10},
  pages={103044},
  year={2023},
  publisher={APS}
}

@article{Vecchiotti_IceCube,
  title={Unveiling the nature of galactic TeV sources with IceCube results},
  author={Vecchiotti, Vittoria and Villante, Francesco L and Pagliaroli, Giulia},
  journal={The Astrophysical Journal Letters},
  volume={956},
  number={2},
  pages={L44},
  year={2023},
  publisher={IOP Publishing}
}

@article{Bustamante_2023,
  title={The Milky Way shines in high-energy neutrinos},
  author={Bustamante, Mauricio},
  journal={Nature Reviews Physics},
  volume={6},
  number={1},
  pages={8--10},
  year={2024},
  publisher={Nature Publishing Group UK London}
}

@article{Pagliaroli,
  title={Expectations for high energy diffuse galactic neutrinos for different cosmic ray distributions},
  author={Pagliaroli, Giulia and Evoli, Carmelo and Villante, Francesco Lorenzo},
  journal={Journal of Cosmology and Astroparticle Physics},
  volume={2016},
  number={11},
  pages={004},
  year={2016},
  publisher={IOP Publishing}
}

@article{Shibata_2010,
  title={A Possible Approach to Three-dimensional Cosmic-ray Propagation in the Galaxy. IV. Electrons and Electron-induced $\gamma$-rays},
  author={Shibata, T and Ishikawa, T and Sekiguchi, S},
  journal={The Astrophysical Journal},
  volume={727},
  number={1},
  pages={38},
  year={2010},
  publisher={IOP Publishing}
}

@article{CR_parameteriation,
  title={Galactic neutrinos in the TeV to PeV range},
  author={Ahlers, Markus and Bai, Yang and Barger, Vernon and Lu, Ran},
  journal={Physical Review D},
  volume={93},
  number={1},
  pages={013009},
  year={2016},
  publisher={APS}
}

@article{ANTARES_Vecchiotti_2023,
  title={Setting an upper limit for the total TeV neutrino flux from the disk of our Galaxy},
  author={Vecchiotti, Vittoria and Villante, Francesco L and Pagliaroli, Giulia},
  journal={Journal of Cosmology and Astroparticle Physics},
  volume={2023},
  number={09},
  pages={027},
  year={2023},
  publisher={IOP Publishing}
}

@article{join_Hawc_IceCube,
  author = {{Alfaro}, R. and {Alvarez}, C. and et al.},
        title = "{Search for Joint Multimessenger Signals from Potential Galactic Cosmic-Ray Accelerators with HAWC and IceCube}",
      journal = {The Astrophysical Journal},
         year = 2024,
        month = nov,
       volume = {976},
       number = {1},
          eid = {8},
        pages = {8},
          doi = {10.3847/1538-4357/ad812f},
archivePrefix = {arXiv},
       eprint = {2405.03817},
 primaryClass = {astro-ph.HE},
       adsurl = {https://ui.adsabs.harvard.edu/abs/2024ApJ...976....8A}
}

@article{ahlera_2014,
  title={Probing the Galactic origin of the IceCube excess with gamma rays},
  author={Ahlers, Markus and Murase, Kohta},
  journal={Physical Review D},
  volume={90},
  number={2},
  pages={023010},
  year={2014},
  publisher={APS}
}

@article{desert_Halzen,
  title={The Milky Way revealed to be a neutrino desert by the IceCube Galactic plane observation},
  author={Fang, Ke and Gallagher, John S and Halzen, Francis},
  journal={Nature Astronomy},
  volume={8},
  number={2},
  pages={241--246},
  year={2024},
  publisher={Nature Publishing Group UK London}
}

@ARTICLE{neutrino_Gagliardini,
       author = {{Gagliardini}, Silvia and {Langella}, Aurora and {Guetta}, Dafne and {Capone}, Antonio},
        title = "{Neutrino Fluxes from Different Classes of Galactic Sources}",
      journal = {The Astrophysical Journal},
         year = 2024,
        month = jul,
       volume = {969},
       number = {2},
          eid = {161},
        pages = {161},
          doi = {10.3847/1538-4357/ad4960},
archivePrefix = {arXiv},
       eprint = {2403.05288},
 primaryClass = {astro-ph.HE},
       adsurl = {https://ui.adsabs.harvard.edu/abs/2024ApJ...969..161G}
}

@article{Steiman-Cameron,
  title={COBE and the galactic interstellar medium: geometry of the spiral arms from FIR cooling lines},
  author={Steiman-Cameron, Thomas Y and Wolfire, Mark and Hollenbach, David},
  journal={The Astrophysical Journal},
  volume={722},
  number={2},
  pages={1460},
  year={2010},
  publisher={IOP Publishing}
}

@ARTICLE{2025PhRvD.111f3035L,
       author = {{Lipari}, Paolo and {Vernetto}, Silvia},
        title = "{Resolved and unresolved Galactic gamma-ray sources}",
      journal = {Physical Review D},
         year = 2025,
        month = mar,
       volume = {111},
       number = {6},
          eid = {063035},
        pages = {063035},
          doi = {10.1103/PhysRevD.111.063035},
archivePrefix = {arXiv},
       eprint = {2412.08861},
 primaryClass = {astro-ph.HE},
       adsurl = {https://ui.adsabs.harvard.edu/abs/2025PhRvD.111f3035L}
}

@article{Abramowski_2014,
   title={Diffuse Galactic gamma-ray emission with H.E.S.S.},
   volume={90},
   ISSN={1550-2368},
   url={http://dx.doi.org/10.1103/PhysRevD.90.122007},
   DOI={10.1103/physrevd.90.122007},
   number={12},
   journal={Physical Review D},
   publisher={American Physical Society (APS)},
   author={{H.~E.~S.~S. Collaboration} and Abramowski and Aharonian et al.},
   year={2014},
   month=dec }

@ARTICLE{2025arXiv250218268D,
       author = {{De La Torre Luque}, Pedro and {Gaggero}, Daniele and {Grasso}, Dario and {Marinelli}, Antonio and {Rocamora}, Manuel},
        title = "{The cosmic-ray sea explains the diffuse Galactic gamma-ray and neutrino emission from GeV to PeV}",
      journal = {arXiv e-prints},
         year = 2025,
        month = feb,
          eid = {arXiv:2502.18268},
        pages = {arXiv:2502.18268},
          doi = {10.48550/arXiv.2502.18268},
archivePrefix = {arXiv},
       eprint = {2502.18268},
 primaryClass = {astro-ph.HE},
       adsurl = {https://ui.adsabs.harvard.edu/abs/2025arXiv250218268D}
}

@ARTICLE{2024PhRvD.109d3007A,
       author = {{Ambrosone}, Antonio and {Groth}, Kathrine M{\o}rch and {Peretti}, Enrico and {Ahlers}, Markus},
        title = "{Galactic diffuse neutrino emission from sources beyond the discovery horizon}",
      journal = {Physical Review D},
         year = 2024,
        month = feb,
       volume = {109},
       number = {4},
          eid = {043007},
        pages = {043007},
          doi = {10.1103/PhysRevD.109.043007},
archivePrefix = {arXiv},
       eprint = {2306.17285},
 primaryClass = {astro-ph.HE},
       adsurl = {https://ui.adsabs.harvard.edu/abs/2024PhRvD.109d3007A}
}

@article{Cao_2023,
   title={Measurement of Ultra-High-Energy Diffuse Gamma-Ray Emission of the Galactic Plane from 10 TeV to 1 PeV with LHAASO-KM2A},
   volume={131},
   ISSN={1079-7114},
   url={http://dx.doi.org/10.1103/PhysRevLett.131.151001},
   DOI={10.1103/physrevlett.131.151001},
   number={15},
   journal={Physical Review Letters},
   publisher={American Physical Society (APS)},
   author={Cao, Zhen and others},
   year={2023},
   month=oct }

@article{Bell,
    author = {Bell, A. R. and Schure, K. M. and Reville, B. and Giacinti, G.},
    title = {Cosmic-ray acceleration and escape from supernova remnants},
    journal = {Monthly Notices of the Royal Astronomical Society},
    volume = {431},
    number = {1},
    pages = {415-429},
    year = {2013},
    month = {02},
    issn = {0035-8711},
    doi = {10.1093/mnras/stt179},
    url = {https://doi.org/10.1093/mnras/stt179},
    eprint = {https://academic.oup.com/mnras/article-pdf/431/1/415/18243010/stt179.pdf},
}

@ARTICLE{2023ApJ...957L...6F,
       author = {{Fang}, Ke and {Murase}, Kohta},
        title = "{Decomposing the Origin of TeV-PeV Emission from the Galactic Plane: Implications of Multimessenger Observations}",
      journal = {\apjl},
         year = 2023,
        month = nov,
       volume = {957},
       number = {1},
          eid = {L6},
        pages = {L6},
          doi = {10.3847/2041-8213/ad012f},
archivePrefix = {arXiv},
       eprint = {2307.02905},
 primaryClass = {astro-ph.HE},
       adsurl = {https://ui.adsabs.harvard.edu/abs/2023ApJ...957L...6F}
}



\end{document}